\documentclass[12pt]{article}
\usepackage{amsfonts,amsmath,amssymb,pifont,amsthm,dsfont,mathtools,mathrsfs}
\usepackage[margin=1.25in]{geometry}
\usepackage{setspace}
\usepackage{graphicx}
\usepackage{appendix}
\usepackage{multirow}
\usepackage{makecell} 
\usepackage{array}
\usepackage{enumerate}
\usepackage{ulem}
\usepackage{cancel}
\usepackage{verbatim}
\usepackage{blkarray}
\usepackage{xcolor}
\usepackage{tikz, tkz-euclide}
\usetikzlibrary{intersections,calc}
\usetikzlibrary{shapes.geometric, arrows, positioning}
\usepackage{caption}
\usepackage{subcaption} 
\usepackage{relsize}
\usepackage{exscale}
\usepackage{booktabs}
\usepackage{longtable}
\usepackage{tabularx} 
\usepackage{ragged2e}

\definecolor{mycolor}{rgb}{0,0,0.75}
\usepackage[round,semicolon]{natbib}
\usepackage[colorlinks, breaklinks,urlcolor=mycolor,citecolor=mycolor,linkcolor=mycolor]{hyperref}
\usepackage{url}

\newtheorem{definition}{Definition}

\newtheorem{property}{Property}

\newtheorem{remark}{Remark}

\newtheorem{proposition}{Proposition}

\begin{document}

	\title{Entropy Regularized Belief Reporting}
	
	\author{Elchin Suleymanov\thanks{Department of Economics, Mitch Daniels Schools of Business, Purdue University, West Lafayette, Indiana, USA. Email: \href{mailto:esuleyma@purdue.edu}{esuleyma@purdue.edu}.}}
	
	\date{\today}
	
	\maketitle
	
	\vspace{-2em}
	\begin{abstract}
		
		This paper investigates a model of partition dependence, a widely reported experimental finding where the agent’s reported beliefs depend on how the states are grouped. In the model, called \textit{Entropy Regularized Belief Reporting~(ERBR)}, the agent is endowed with \textit{a latent benchmark prior} that is unobserved by the analyst. When presented with a partition, the agent reports a prior that minimizes Kullback-Leibler divergence from the latent benchmark prior subject to entropy regularization. This captures the intuition that while the agent would like to report a prior that is close to her latent benchmark prior, she may also have a preference to remain noncommittal. I provide the structural properties of the model that allow for identification of the latent benchmark prior and apply the model to the experimental data from \citet{benjamin2017biased}. 
		
		\medskip
		
				
		
	\end{abstract}
	
	\thispagestyle{empty}
	\clearpage
	
	\setcounter{page}{1}
	\pagestyle{plain}
	
	\section{Introduction}
	
	Consider the following two thought experiments. In the first, subjects are asked to evaluate the probabilities for the events that the S\&P 500 will finish the year below 7,000 or above 7,000. In the second, subjects are asked to evaluate the probabilities for the events that the S\&P 500 will finish the year below 6,500, between 6,500 and 7,000, or above 7,000. A consistent result within the experimental literature demonstrates that we should expect subjects to assign a much higher probability to the event that the S\&P 500 finishes the year above 7,000 in the first thought experiment. This phenomenon, known as partition dependence, was formalized by \citet{tversky1994support} building on previous experimental literature on the topic, and has since been demonstrated in numerous experiments using both direct belief elicitation and incentivized choice data \citep[for a recent review, see][]{benjamin2019errors}. 

	From an analyst's perspective, partition dependence has significant implications for belief elicitation in experiments, the measurement of biases in probabilistic reasoning where partition dependence can act as a confounder, and the design of public surveys. This motivates a model that allows the analyst to separate potentially biased elicited beliefs from the agent’s latent beliefs. From a firm's perspective, partition dependence can potentially be exploited to charge a higher price for an inferior product. For example, in an experiment, \citet{johnson1993framing} show that many decision-makers are willing to pay a higher amount for insurance that covers hospitalization for any disease or accident than for insurance that covers hospitalization for any reason, despite the fact that the latter insurance includes the former. Hence, if a firm can infer agents’ latent beliefs from observed data, it can use this information in product design to improve profits.

	Partition dependence has also been shown to affect practitioners, raising questions about the validity of probability estimates one might receive from experts. For example, partition dependence has been observed among professional forecasters \citep{sonnemann2013psychological}, decision analysts \citep{fox2005subjective}, attorneys \citep{fox2002forecasting}, options traders \citep{fox1996options}, medical professionals \citep{redelmeier1995probability}, restaurant managers \citep{dube1988availability}, and auto mechanics \citep{fischhoff1978fault}. The finding in \citet{fox1996options} is especially notable as it shows that options traders' choices are consistent with expected utility in the risk domain while exhibiting partition dependence in the uncertainty domain.
	
	Given the prevalence of partition dependence and its significant implications, there is a need for a simple and tractable model that can be employed by analysts. Building on recent variational frameworks that model belief distortions via entropy regularization \citep[e.g.,][]{prat2021biases, little2022information, strzalecki2024variational}, I explore one such model and investigate its implications. In the model, called \textit{Entropy Regularized Belief Reporting~(ERBR)}, the agent is endowed with a latent benchmark prior $\pi$ over the state space $\Omega$ that is unobserved by the analyst. The agent's reported beliefs do not necessarily reflect her latent prior and are sensitive to the partitioning of the state space. Formally, given a partition $\mathcal{P}$ of the state space, the agent reports a belief $\mu_{\mathcal{P}} \in \Delta(\mathcal{P})$. The belief the agent reports balances two concerns. First, the agent wants to report beliefs that are close to those induced by her latent benchmark. Second, the agent also has a preference to remain noncommittal, captured by a desire for higher entropy in the reported beliefs. Letting $\pi_{\mathcal{P}}\in \Delta(\mathcal{P})$ be such that $\pi_{\mathcal{P}}(E_i) = \pi(E_i)$ for each $E_i\in \mathcal{P}$, the agent's reported belief satisfies
	\begin{equation*}
		\mu_{\mathcal{P}} = \underset{\pi'_{\mathcal{P}}\in \Delta(\mathcal{P})}{\arg\min} \:\: \lambda D_{KL}\left(\pi'_{\mathcal{P}} || \pi_{\mathcal{P}}\right)- (1-\lambda) H\left(\pi'_{\mathcal{P}}\right),
	\end{equation*}
	where $D_{KL}(\cdot||\cdot)$ denotes Kullback-Leibler divergence, $H(\cdot)$ denotes entropy, and $\lambda$ is a parameter weighting these two concerns. While the base case in the model is when $\lambda\in [0,1]$, the objective is well-defined for any $\lambda \in \mathbb{R}$. We can interpret $\lambda<0$ as a preference for misreporting and $\lambda>1$ as a preference for extreme beliefs or overconfidence.
	
	The optimization problem given above leads to a belief reporting rule
	\begin{equation*}
	\mu_\mathcal{P}(E_i) = \frac{\pi(E_i)^{\lambda}}{\sum_{E_j\in \mathcal{P}} \pi(E_j)^{\lambda}}.
	\end{equation*}
	When $\lambda = 1$, the agent reports her beliefs accurately. On the other hand, when $\lambda<1$, the reported beliefs satisfy the property known as subadditivity. In the context of the thought experiment, this means that the subjective probability assigned to the event ``the S\&P 500 finishes below 7,000'' is lower when it is presented as a single event than when it is unpacked into constituent sub-events. Conversely, the agent assigns a higher probability to the event that the S\&P 500 finishes above 7,000 in the binary partition compared to the ternary partition. To elaborate, let $E_1$ = \{S\&P finishes above 7,000\}, $E_2$ = \{S\&P finishes between 6,500 and 7,000\}, and $E_3$ = \{S\&P finishes below 6,500\}. Let $\pi(E_1) = p$ and $\pi(E_2) = q$, so that $\pi(E_3) = 1-p-q$. Let $\mathcal{P}_1 = \{E_1, E_2\cup E_3\}$ denote the binary partition and $\mathcal{P}_2 = \{E_1, E_2, E_3\}$ be the ternary partition. We then have
	\begin{equation*}
		\mu_{\mathcal{P}_1}(E_1) = \frac{p^{\lambda}}{p^{\lambda}+(1-p)^{\lambda}} \quad \text{and} \quad  \mu_{\mathcal{P}_2}(E_1) = \frac{p^{\lambda}}{p^{\lambda}+q^{\lambda}+(1-p-q)^{\lambda}}.
	\end{equation*}
	When $p+q<1$ and $\lambda<1$, this implies $\mu_{\mathcal{P}_1}(E_1)>\mu_{\mathcal{P}_2}(E_1)$, consistent with most experimental literature. When $\lambda>1$, the agent reports extreme beliefs that satisfy the property known as superadditivity. While most existing evidence is consistent with subadditivity, superadditivity has also been reported in the experimental literature \citep{macchi1999note, sloman2004typical}.
	
	In Section \ref{sec:model}, I derive the structural properties of the model that are necessary for identification of latent beliefs and apply the model to the experimental data from \citet{benjamin2017biased}. I show that while the model significantly improves the fit relative to the rational benchmark ($\lambda=1$), the improvement is not uniform across all partitions. In particular, for some partitions, the rational benchmark itself provides a better fit. This strongly suggests that the entropy regularization parameter, $\lambda$, is likely partition-dependent. This contrasts with previous variational literature \citep{prat2021biases, little2022information, strzalecki2024variational}, which assumes that the entropy regularization parameter is constant across events and partitions. The intuition for why this parameter might be partition-dependent is as follows. While the agent has a benchmark belief over the entire state space, her confidence in evaluating the specific event probabilities may vary depending on the given partition. For example, the agent might have a highly confident assessment that the S\&P finishes the year below 7,000, but her confidence in her assessments for the sub-events ``between 6,500 and 7,000'' and ``below 6,500'' may not be as high. Since the agent's confidence can vary across partitions, this leads to a partition-dependent regularization parameter.

	This extension modifies the belief reporting rule to
	\begin{equation*}
		\mu_\mathcal{P}(E_i) = \frac{\pi(E_i)^{\lambda_{\mathcal{P}}}}{\sum_{E_j\in \mathcal{P}} \pi(E_j)^{\lambda_{\mathcal{P}}}}.
	\end{equation*}
	In Section \ref{sec:partitiondep}, I show that the version of the model with partition-dependent parameters matches the experimental data in \citet{benjamin2017biased} quite well. Importantly, the improvement remains if one assumes constant entropy regularization for binary partitions. This allows me to recover the subjects' latent beliefs from binary partitions and evaluate the performance of the model using the recovered latent beliefs rather than the objective coin-flipping prior implied by the experimental design. I show that using recovered latent beliefs improves the out-of-sample performance in other partitions and results in recovered entropy regularization parameters that are closer to one. This suggests that the recovered beliefs are a better reflection of the subjects' underlying beliefs than the objective prior in the experiment. Lastly, I also show that the extended version of the model can accommodate counterintuitive, yet widely observed fallacies such as the conjunction and disjunction effects reported in \citet{tversky1983extensional} and \citet{barhillel1993how}.

	\subsection{Related Literature}
	\label{sec:lit}
	
	The leading theory of partition dependence is known as Support Theory, which was developed in \citet{tversky1994support} and \citet{rottenstreich1997unpacking} and later axiomatized in \citet{ahn2010framing}. Support Theory posits the existence of a non-negative support function $s(\cdot)$ that is defined for all events. The reported beliefs in Support Theory are given by 
	\begin{equation*}
		\label{eq:rep_beliefs_support}
		\mu_\mathcal{P}(E_i) = \frac{s(E_i)}{\sum_{E_j\in \mathcal{P}} s(E_j)}.
	\end{equation*}
	\citet{tversky1994support} also assume that the support function is subadditive: $s(E_i\cup E_j) \leq s(E_i)+s(E_j)$ for $E_i$ and $E_j$ with an empty intersection. This generates subadditivity in the reported beliefs. Letting $s(E) = \pi(E)^{\lambda}$, the ERBR model with a constant entropy regularization parameter $\lambda\leq 1$ becomes a special case of Support Theory. Hence, we can view the single-parameter ERBR model as providing a special, more tractable version of Support Theory. In addition, the ERBR model allows the analyst to recover the latent benchmark probabilities of the agent from the observed beliefs, as demonstrated in Sections \ref{sec:model} and \ref{sec:partitiondep}. Furthermore, the extended model with partition-dependent parameters generalizes Support Theory by relaxing the assumption that the support for an event is invariant to the partition structure.
	
	 There have been a few attempts to extend Support Theory to better match observed experimental data. \citet{fox2003partition} and \citet{see2006between} extend Support Theory by proposing that the reported belief for an event depends both on the support for that event and the ignorance prior that assigns equal probabilities to all partition elements. \citet{clemen2008interior} propose an alternative approach where the reported belief is a convex combination of the latent belief and the uniform prior over partition elements. There is also an extensive psychology literature documenting and modeling partition dependence \citep[for a recent review, see][]{huang2025models}. As I discuss in Section \ref{sec:model}, the ERBR model provides a theoretical microfoundation for the approaches relying on the ignorance prior to explain partition dependence: the preference for entropy in the objective function is mathematically equivalent to a preference for minimizing divergence from the ignorance prior. This allows the ERBR model to capture the same empirical regularities within a tractable variational framework that is common in economic models.
	
	\citet{fox2005partition} argue that partition dependence in belief reporting is just one manifestation of a broad phenomenon where agents use a ``maximum entropy heuristic.'' In their review of experimental studies across various domains such as decision analysis, organizational resource allocation, and consumer choice, they show that reliance on this heuristic can systematically explain deviations from the rational benchmark. They conclude that subjects typically make choices based on a combination of the maximum entropy heuristic and what would be dictated by subjective preferences and beliefs. Similarly, in the ERBR model, subjects balance the preference for closeness of their reported beliefs to their latent benchmark with the preference for entropy.
	
	The idea of entropy regularization has also been used in the literature to explain deviations from Bayesian updating. \citet{prat2021biases} use a similar objective function where the agent chooses a posterior that minimizes Kullback-Leibler divergence from the Bayesian posterior plus an additive term that captures entropy regularization. \citet{little2022information} and \citet{strzalecki2024variational} extend this idea to derive the exponential updating rule in \citet{grether1980bayes}, a widely used model that captures most commonly observed updating biases. \citet{little2022information} also shows that the same objective function can be used as a unified framework to accommodate various other biases in probabilistic reasoning, including partition dependence. I build upon the analysis in these papers in three specific ways. First, I provide the structural properties of the model with constant entropy regularization that can be used to test the model. Second, I show that while the model with constant entropy regularization relied upon in the previous papers improves upon the rational benchmark in the aggregate, the improvement is partition-dependent and may produce inferior performance for some partitions. Lastly, I extend the model to allow for partition-dependent entropy regularization, which significantly improves the model performance, and demonstrate how the model can be applied to recover latent beliefs from observed data. Crucially, recovering latent beliefs helps disentangle biases in belief reporting from other biases in probabilistic reasoning.

	\section{The Model}
	\label{sec:model}
	
	Let $\Omega$ be a finite set of states and $\Delta(\Omega)$ be the set of probability distributions on $\Omega$. An event $E$ is a subset of $\Omega$. The agent has a \textit{benchmark prior} $\pi\in \Delta(\Omega)$, assumed to have full support, that an analyst would like to elicit. In particular, the analyst would like to infer the probability of $E$ for each $E\subseteq \Omega$. To this end, the analyst presents the agent with a partition of the state space $\mathcal{P}$. A partition is a collection of non-empty, disjoint events $\{E_1, E_2, \ldots, E_k\}$ whose union is $\Omega$, for some $k\in \{1,\ldots, |\Omega|\}$. When the analyst presents the agent with the partition $\mathcal{P}$, she wants to elicit the probability of each event $E_i$ for $E_i\in \mathcal{P}$. The collection of all partitions of the state space is denoted by $\mathbb{P}$.

	When faced with the partition $\mathcal{P}$, the agent reports a belief $\mu_{\mathcal{P}}\in \Delta(\mathcal{P})$. The reported belief $\mu_{\mathcal{P}}$ can be interpreted as the agent's observable partition-dependent probabilistic judgment, which can be elicited either directly via a survey or indirectly from incentivized choice behavior. As discussed in the Introduction, partition dependence has been documented in both contexts. Note that $\mu_{\mathcal{P}}$ only assigns beliefs to events $E_i\in \mathcal{P}$. Let $\pi_{\mathcal{P}}$ denote the element in $\Delta(\mathcal{P})$ induced by the benchmark prior $\pi$, i.e., $\pi_{\mathcal{P}}(E_i) = \pi(E_i)$ for all $E_i\in \mathcal{P}$. If the agent reports her latent benchmark prior accurately, then $\mu_{\mathcal{P}}(E_i)=\pi_{\mathcal{P}}(E_i)$ for all $E_i\in \mathcal{P}$. However, for an agent with partition-dependent reported beliefs, $\mu_{\mathcal{P}}$ will generally be distinct from $\pi_{\mathcal{P}}$. This poses a problem for the analyst's belief elicitation exercise. 
	
	I next present a model that allows the analyst to recover the latent benchmark prior $\pi$ from reported beliefs $\{\mu_{\mathcal{P}}\}_{\mathcal{P}\in\mathbb{P}}$. In the model, called \textit{Entropy Regularized Belief Reporting~(ERBR)}, when faced with the partition $\mathcal{P}$, the agent reports a belief $\mu_{\mathcal{P}}$ that balances the preference for reporting beliefs that are close to those induced by the benchmark prior, where closeness is defined by Kullback-Leibler (KL) divergence, and the preference for being noncommittal, which is captured by entropy. The next definition formally describes the model.
	
	\begin{definition}
		\label{def:model}
		A belief reporting rule $\{\mu_{\mathcal{P}}\}_{\mathcal{P}\in\mathbb{P}}$ is consistent with the \textit{Entropy Regularized Belief Reporting~(ERBR)} model if there exist a benchmark prior $\pi\in \Delta(\Omega)$, assumed to have full support, and an entropy regularization parameter $\lambda\in \mathbb{R}$ such that
		$$\mu_{\mathcal{P}} = \underset{\pi'_{\mathcal{P}}\in \Delta(\mathcal{P})}{\arg\min} \:\: \lambda D_{KL}\left(\pi'_{\mathcal{P}} || \pi_{\mathcal{P}}\right)- (1-\lambda) H\left(\pi'_{\mathcal{P}}\right),$$
		where 
		$$D_{KL}(\pi'_{\mathcal{P}} || \pi_{\mathcal{P}} ) = \sum_{E_i \in \mathcal{P}}\pi'_{\mathcal{P}}(E_i) \ln\left(\frac{\pi'_{\mathcal{P}}(E_i)}{\pi_{\mathcal{P}}(E_i)}\right)  \:\: \text{and}  \:\: H\left(\pi'_{\mathcal{P}}\right) = -\sum_{E_i \in \mathcal{P}}\pi'_{\mathcal{P}}(E_i) \ln\left(\pi'_{\mathcal{P}}(E_i)\right).$$
	\end{definition}
	
	To illustrate the model, consider the following cases. First, if $\lambda=1$, then the agent always reports the probabilities implied by her benchmark prior. This reflects the standard case where the analyst can easily elicit the agent's beliefs. Alternatively, if $\lambda=0$, the agent reports beliefs that maximize the entropy within $\Delta(\mathcal{P})$. Since the uniform prior in $\Delta(\mathcal{P})$, denoted by $u_{\mathcal{P}}$, maximizes the entropy, this will be the reported belief in this case. This corresponds to an agent who reports beliefs that are maximally noncommittal, and the analyst can no longer elicit the agent's latent beliefs in this case. If $\lambda\in (0,1)$, which is most consistent with the experimental literature, the agent reports a belief that balances the two concerns in the model.
	
	While the base case in the model is when $\lambda\in [0,1]$, the objective in Definition \ref{def:model} is well-defined for any $\lambda\in \mathbb{R}$. We can interpret the cases when $\lambda \notin [0,1]$ as follows. If $\lambda<0$, then the agent has a preference for beliefs that are further away from her benchmark prior. This captures a preference for misreporting. Alternatively, if $\lambda>1$, then the agent has a preference for lower entropy in the reported beliefs. This captures an agent who is overconfident and reports more extreme beliefs. 
	
	\begin{remark}
		\label{remark:wdm}
		We can alternatively write the model as 
		$$\mu_{\mathcal{P}} = \underset{\pi'_{\mathcal{P}}\in \Delta(\mathcal{P})}{\arg\min} \:\: \lambda D_{KL}\left(\pi'_{\mathcal{P}} || \pi_{\mathcal{P}}\right)+ (1-\lambda)D_{KL}\left(\pi'_{\mathcal{P}} || u_{\mathcal{P}}\right),$$
		where $u_{\mathcal{P}}$ is the uniform prior on $\mathcal{P}$. This follows from the fact that
		\begin{align*}
			D_{KL}\left(\pi'_{\mathcal{P}} || u_{\mathcal{P}}\right) &=  \sum_{E_i \in \mathcal{P}}\pi'_{\mathcal{P}}(E_i) \ln\left(\frac{\pi'_{\mathcal{P}}(E_i)}{u_{\mathcal{P}}(E_i)}\right) \\
			& =  \sum_{E_i \in \mathcal{P}}\pi'_{\mathcal{P}}(E_i) \ln\left(\pi'_{\mathcal{P}}(E_i)\right) -  \sum_{E_i \in \mathcal{P}}\pi'_{\mathcal{P}}(E_i) \ln\left(u_{\mathcal{P}}(E_i)\right)\\
			& = -H(\pi'_{\mathcal{P}}) - \sum_{E_i \in \mathcal{P}}\pi'_{\mathcal{P}}(E_i) \ln\left(\frac{1}{|\mathcal{P}|}\right) \\
			& = -H(\pi'_{\mathcal{P}})  + \ln(|\mathcal{P}|).
		\end{align*}
		In this alternative formulation, the agent minimizes weighted KL divergences from her benchmark prior and the uniform prior on $\mathcal{P}$. This provides a microfoundation for existing approaches in the literature that model partition dependence as a pull towards the ignorance prior \citep[see][]{fox2003partition, see2006between, clemen2008interior}. For $\lambda\in [0,1]$, the weights for both KL divergences are non-negative. For $\lambda\notin [0,1]$, the coefficient for one of the KL divergences becomes negative. 
	\end{remark}
	
	The next proposition provides a formula connecting observed partition-dependent beliefs to the benchmark prior and the entropy regularization parameter. 
	
	\begin{proposition}
		\label{prop:exponential}
		Suppose the observed partition-dependent beliefs $\{\mu_{\mathcal{P}}\}_{\mathcal{P}\in \mathbb{P}}$ are consistent with the ERBR model with the benchmark prior $\pi$ and the entropy regularization parameter $\lambda$. Then, for any $\mathcal{P}\in \mathbb{P}$ and $E_i\in \mathcal{P}$,
		\begin{equation}
			\label{eq:reported_beliefs}
			\mu_\mathcal{P}(E_i) = \frac{\pi_\mathcal{P}(E_i)^{\lambda}}{\sum_{E_j\in \mathcal{P}} \pi_\mathcal{P}(E_j)^{\lambda}} = \frac{\pi(E_i)^{\lambda}}{\sum_{E_j\in \mathcal{P}} \pi(E_j)^{\lambda}} .
		\end{equation}
	\end{proposition}
	
	If $\lambda \leq 1 $, then the reported beliefs induced by the ERBR model are consistent with Support Theory developed in \citet{tversky1994support} and extended in \citet{rottenstreich1997unpacking}. In Support Theory, the agent is endowed with a support function $s: 2^{\Omega}\setminus \{\emptyset\} \rightarrow \mathbb{R}_{++}$ such that
	$$\mu_\mathcal{P}(E_i) = \frac{s(E_i)}{\sum_{E_j\in \mathcal{P}} s(E_j)}.$$
	\citet{tversky1994support} assume a subadditive support function, i.e., $s(E_i\cup E_j)\leq s(E_i)+s(E_j)$ for $E_i,E_j\subseteq \Omega$ with $E_i\cap E_j=\emptyset$. 
	The next proposition formally shows that if we let $s(E_i) = \pi(E_i)^\lambda,$ then subadditivity holds as long as $\lambda\leq 1$. When $\lambda\geq 1$, the support function becomes superadditive: $s(E_i\cup E_j)\geq s(E_i)+s(E_j).$
	
	\begin{proposition}
		\label{prop:subadditive}
		Let $s: 2^{\Omega}\setminus \{\emptyset\} \rightarrow \mathbb{R}_{++}$ be a support function defined by $s(E) = \pi(E)^{\lambda}$, where $\pi$ is the full support benchmark prior and $\lambda\in \mathbb{R}$. Then, 
		\begin{itemize}
			\item $s(\cdot)$ is subadditive for $\lambda\leq 1$,
			\item $s(\cdot)$ is superadditive for $\lambda\geq 1$. 
		\end{itemize}
	\end{proposition}
	
	Reported beliefs are said to be consistent with Generalized Support Theory (GST) if the support function is allowed to be subadditive or superadditive. The ERBR model is a subclass of the GST model.
	
	\begin{definition}
		\label{def:support}
		A belief reporting rule $\{\mu_{\mathcal{P}}\}_{\mathcal{P}\in\mathbb{P}}$ is consistent with \textit{Generalized Support Theory} (GST) if there exists a support function $s: 2^{\Omega}\setminus \{\emptyset\} \rightarrow \mathbb{R}_{++}$ such that
		\begin{equation}
			\label{eq:GST}
			\mu_\mathcal{P}(E_i) = \frac{s(E_i)}{\sum_{E_j\in \mathcal{P}} s(E_j)}
		\end{equation}
		for all $\mathcal{P}\in \mathbb{P}$ and $E_i\in \mathcal{P}$. 
	\end{definition}
	
	\subsection{Structural Properties and Identification of Latent Beliefs}
	\label{sec:properties}
	
	In this section, I establish the structural properties required to identify the latent benchmark prior from reported beliefs. To this end, I take the agent's partition-dependent reported beliefs $\{\mu_{\mathcal{P}}\}_{\mathcal{P}\in \mathbb{P}}$ as the observable primitive of the analysis. I first show that two properties, Regularity and Cyclical Independence, characterize the GST model. I then show that if the support function identified by these properties satisfies Power Additivity, then the beliefs must be consistent with the ERBR model.
	
	The first property, Regularity, is a standard technical condition requiring that for any given partition, reported beliefs are strictly positive and sum to one. The ERBR model satisfies positivity due to the full support assumption on the latent prior $\pi$, and the GST model satisfies it due to the assumption that $s(A)>0$ for all nonempty $A\subseteq \Omega$.
	
	\begin{property}[Regularity]
		\label{property:regularity}
		For any $\mathcal{P}\in \mathbb{P}$ and $E_i\in \mathcal{P}$, $\mu_{\mathcal{P}}(E_i)>0$. In addition, $\sum_{E\in \mathcal{P}}\mu_{\mathcal{P}}(E)=1$. 
	\end{property}
	
	In most experimental studies, when the subjects are given a partition $\mathcal{P}$, they are instructed that the probabilities for the events in the given partition must sum to one. As the experiment in \citet{teigen1974subjective} demonstrates, when the subjects are not explicitly given this instruction, the sum of probabilities for the events in the partition tends to be greater than one. This likely reflects the intuition that if the subjects are not given the specific instruction, they evaluate the probability for each event in the partition separately, by forming the partition $\mathcal{P}_{E_i}=\{E_i, E_i^c\}$ for each $E_i\in \mathcal{P}$, and reporting $\mu_{\mathcal{P}_{E_i}}(E_i)$ rather than $\mu_{\mathcal{P}}(E_i)$. We would typically expect $\sum_{E_i \in \mathcal{P}}\mu_{\mathcal{P}_{E_i}}(E_i)>1$ under subadditivity, coinciding with $\lambda < 1$ in the model. Hence, in experimental studies, it is important to specify that the probabilities for the events in the partition $\mathcal{P}$ should sum to one to ensure that the subjects report the appropriate partition-dependent belief $\mu_{\mathcal{P}}$ and not $\mu_{\mathcal{P}_{E_i}}$.
	
	To introduce the next property, consider two partitions $\mathcal{P}_1$, $\mathcal{P}_2$ and two events $E_1,E_2\in \mathcal{P}_1\cap \mathcal{P}_2$. The representation for the GST model given by equation \ref{eq:GST} implies that
	$$\frac{\mu_{\mathcal{P}_1}(E_1)}{\mu_{\mathcal{P}_1}(E_2)} = \frac{s(E_1)}{s(E_2)} = \frac{\mu_{\mathcal{P}_2}(E_1)}{\mu_{\mathcal{P}_2}(E_2)} \quad \Rightarrow \quad \frac{\mu_{\mathcal{P}_1}(E_1)}{\mu_{\mathcal{P}_1}(E_2)}\frac{\mu_{\mathcal{P}_2}(E_2)}{\mu_{\mathcal{P}_2}(E_1)}=1.$$
	The next property, Cyclical Independence, generalizes this idea for arbitrary cycles of partitions and events.
	
	\begin{property}[Cyclical Independence]
		\label{property:cyclical}
		For any collection of partitions $\mathcal{P}_1, \mathcal{P}_2, \ldots, \mathcal{P}_n$ and any collection of events $E_1,E_2,\ldots, E_n$ that satisfies $E_{i+1}\in \mathcal{P}_i\cap \mathcal{P}_{i+1}$ for $i<n$ and $E_1\in \mathcal{P}_1 \cap \mathcal{P}_n$, we have 
		$$\frac{\mu_{\mathcal{P}_1}(E_1)}{\mu_{\mathcal{P}_1}(E_2)}\frac{\mu_{\mathcal{P}_2}(E_2)}{\mu_{\mathcal{P}_2}(E_3)}\cdots \frac{\mu_{\mathcal{P}_n}(E_n)}{\mu_{\mathcal{P}_n}(E_1)}=1.$$
	\end{property}
	
	As the next proposition shows, Regularity and Cyclical Independence are necessary and sufficient for the GST model. A characterization of the GST model was provided by \citet{ahn2010framing}, who take partition-dependent preferences $\{\succeq_{\mathcal{P}}\}_{\mathcal{P}\in\mathbb{P}}$ over \citet{anscombe1963definition} acts as the primitive of the analysis. If we observe $\{\succeq_{\mathcal{P}}\}_{\mathcal{P}\in\mathbb{P}}$ as in \citet{ahn2010framing}, then we can elicit partition-dependent beliefs $\{\mu_{\mathcal{P}}\}_{\mathcal{P}\in\mathbb{P}}$ from these preferences. Alternatively, analysts often observe beliefs directly (e.g., probability forecasts or direct belief elicitation in experiments) without access to the underlying preference data. The next proposition provides a characterization and an identification result for the GST model using this simpler, more commonly observed primitive.
	
	\begin{proposition}
		\label{prop:GST}
		A collection of partition-dependent beliefs $\{\mu_{\mathcal{P}}\}_{\mathcal{P}\in \mathbb{P}}$  is consistent with Generalized Support Theory if and only if it satisfies Regularity and Cyclical Independence. In addition, the support function $s(\cdot)$ is uniquely identified for all proper events up to scale normalization. 
	\end{proposition}
	
	From the previous proposition, the support function $s(\cdot)$ is uniquely identified for all proper events up to scale normalization. Importantly, this identification is constructive: it provides a concrete algorithm for the analyst to recover the support values from observed reported beliefs. To illustrate, fix a state $\omega^*\in \Omega$ and let $s(\omega^*)=1$. Consider a proper event $A\subsetneq \Omega$. If $\omega^*\notin A$, then consider the partition $\mathcal{P}= \{A, A^c\setminus \{\omega^*\},  \{\omega^*\}\}$ and define $s(A)$ by
	$$s(A) = \frac{\mu_{\mathcal{P}}(A)}{\mu_{\mathcal{P}}(\{\omega^*\})}.$$
	Alternatively, if $\omega^*\in A$, then pick $\omega\notin A$. Since $A\subsetneq \Omega$, such an $\omega$ exists. Construct $\mathcal{P}_1 = \{\{\omega^*\}, \{\omega\} , \{\omega^*, \omega\}^c\}$, $\mathcal{P}_2 = \{A, A^c\setminus \{\omega\}, \{\omega\}\}$ and let
	$$s(A) = \frac{\mu_{\mathcal{P}_2}(A)}{\mu_{\mathcal{P}_2}(\{\omega\})} \frac{\mu_{\mathcal{P}_1}(\{\omega\})}{\mu_{\mathcal{P}_1}(\{\omega^*\})}.$$
	Hence, using at most two partitions, the analyst can uniquely identify $s(\cdot)$ up to normalization.
	
	Let $s(\cdot)$ be a support function identified from $\{\mu_{\mathcal{P}}\}_{\mathcal{P}\in \mathbb{P}}$. The next property requires that $s(\cdot)^{\alpha}$ is additive for some $\alpha\neq 0$. 
	
	\begin{property}[Power Additivity]
		\label{property:additivity}
		Either $s(\cdot)$ is constant or there exists $\alpha \neq 0$ such that for any $A,B\subsetneq \Omega$ with $A\cap B=\emptyset$ and $A\cup B\subsetneq \Omega$,
		$$s(A\cup B)^{\alpha} = s(A)^{\alpha}+s(B)^{\alpha}.$$
	\end{property}
	
	Since the ERBR model is a special case of the GST model with $s(A) = \pi(A)^\lambda,$ the necessity of the property is clear by letting $\alpha = 1/\lambda$ whenever $\lambda\neq 0$ so that $s(\cdot)$ is not constant. While Property \ref{property:additivity} is an existence property, testing it is quite straightforward. Once the support function is identified through the construction defined above, pick any two events $A$ and $B$ with $A\cap B=\emptyset$ and $A\cup B\subsetneq \Omega$ (e.g., $A$ and $B$ can be two singleton events). Solve the equation 
	$$s(A\cup B)^{\alpha} = s(A)^{\alpha}+s(B)^{\alpha}$$
	for some $\alpha$. If no solution exists, then the property is automatically violated. If a solution does exist, it will have to be unique. In fact, if $\min\{s(A),s(B)\}\leq s(A\cup B)\leq \max\{s(A),s(B)\}$, then we can show that no solution to the above equation can exist. Alternatively, if $s(A\cup B)> \max\{s(A),s(B)\}$, then there will always be a unique solution with $\alpha>0$, and if $s(A\cup B)< \min\{s(A),s(B)\}$, then there will always be a unique solution with $\alpha<0$ for any pair $A$ and $B$.\footnote{To see this, let $s(A)=a, s(B)=b$, and $s(A\cup B)=c$. First, assume $b\leq c\leq a$ (the case when $a\leq c\leq b$ is analogous). Then, $a^{\alpha}\geq c^{\alpha}$ whenever $\alpha>0$ and $b^{\alpha}\geq c^{\alpha}$ whenever $\alpha<0$. Since $a,b>0$, the equation $c^{\alpha}=a^{\alpha}+b^{\alpha}$ cannot hold for any $\alpha$. Alternatively, suppose $c>\max\{a,b\}$. Let $g(\alpha) = (\frac{a}{c})^{\alpha}+(\frac{b}{c})^{\alpha}$. Since $g$ is continuous, strictly decreasing with $g(0)=2$ and $\lim_{\alpha\rightarrow \infty} g(\alpha) = 0$, $g(\alpha)=1$ for a unique $\alpha>0$. The case when $c<\min\{a,b\}$ is similar.} Hence, by considering two events, we will either deduce that the property is violated or get a unique candidate $\alpha$ for the property. We can then construct $s(\cdot)^{\alpha}$ and verify if additivity holds.
	
	The next proposition provides a characterization and an identification result for the ERBR model and shows that the benchmark prior is uniquely recovered unless $\{\mu_{\mathcal{P}}\}_{\mathcal{P}\in \mathbb{P}}$ is always uniform. 
	
	\begin{proposition}
		\label{prop:ERBR}
		Suppose the collection of partition-dependent beliefs $\{\mu_{\mathcal{P}}\}_{\mathcal{P}\in \mathbb{P}}$ admits a Generalized Support Theory representation with a support function $s(\cdot)$. Then, the reported beliefs are represented by the Entropy Regularized Belief Reporting (ERBR) model if and only if $s(\cdot)$ satisfies Property \ref{property:additivity}.
		
		In addition, the entropy regularization parameter $\lambda$ is uniquely identified and the benchmark prior $\pi$ is uniquely recovered unless $\{\mu_{\mathcal{P}}\}_{\mathcal{P}\in \mathbb{P}}$ is always uniform.
	\end{proposition}
	
	\subsection{Empirical Investigation}
	\label{sec:empirics}
	
	In this section, I investigate how the model aligns with the experimental evidence from \citet{benjamin2017biased}. In their experiment, a fair coin was tossed ten times, and participants were asked to provide their probability estimates for the number of resulting heads. They elicited beliefs for various partitions, as defined in Table \ref{tab:benjamin}.
	
	\begin{table}[htbp]
		\centering
		\begin{tabularx}{\linewidth}{l >{\RaggedRight}X}
			\toprule
			\textbf{Partition} & \textbf{Events (Bins)} \\
			\midrule
			$\mathcal{P}_1$ & $\{\{0\}, \{1\}, \ldots, \{10\}\}$ \\
			$\mathcal{P}_2$ & $\{\{0,1,2,3\}, \{4\}, \{5\}, \{6\}, \{7,8,9,10\}\}$ \\
			$\mathcal{P}_3$ & $\{\{0,1,2,3,4\}, \{5\}, \{6, 7,8,9,10\}\}$ \\
			$\mathcal{P}_{4,k}$ & $\{\{k\}, \{k\}^c\}$ for each $k \in \{0,\ldots, 10\}$ \\
			\bottomrule
		\end{tabularx}
		\caption{Partitions used in \citet{benjamin2017biased}.}
		\label{tab:benjamin}
	\end{table}
	
	They conduct two experiments: Experiment 1 with 104 customers at a food court in Pittsburgh, Pennsylvania, and a follow-up, Experiment 2, with 308 UC Berkeley students (see \cite{benjamin2017biased} for full details). I use the data from Experiment 2 for the subsequent analysis, as its subjects were significantly more probabilistically sophisticated.\footnote{The analysis of data from Experiment 1 results in a best-fitting entropy regularization parameter of $\lambda \approx 0.23$, suggesting most subjects defaulted to a uniform distribution to a large extent. In contrast, the best-fitting parameter for Experiment 2 was $\lambda\approx 0.69$.} The empirical data for this paper are taken from Appendix J, Figures 9-12, posted on the Open Science Framework at \href{https://osf.io/mjwbi/}{https://osf.io/mjwbi/}.
	
\begin{figure}[htbp]
		\centering
		\includegraphics[width=\linewidth]{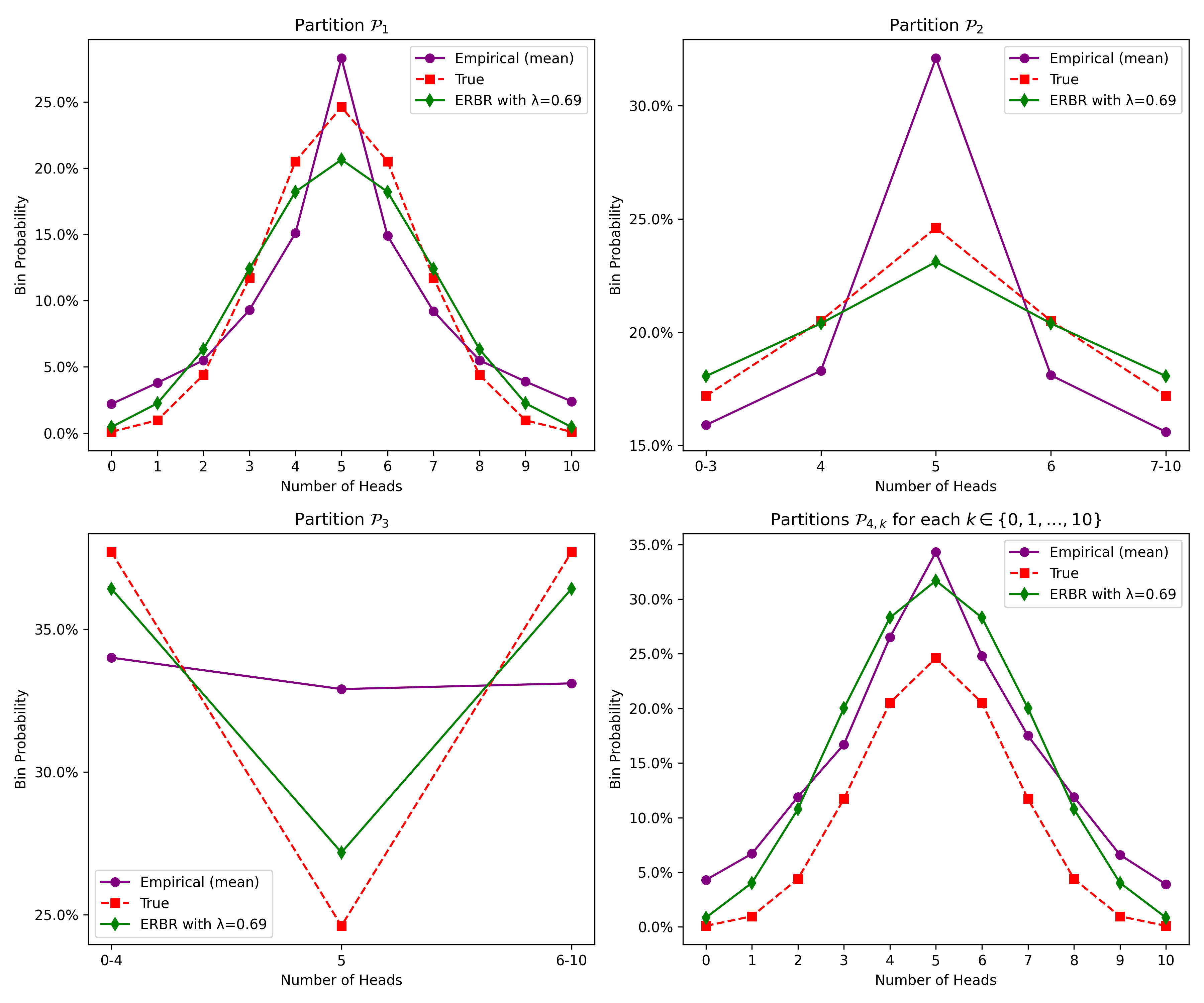} 
		\caption{Event (or bin) probabilities for each partition. Empirical data (purple) reflects the mean probabilities from Experiment 2 in \cite{benjamin2017biased}. True probabilities are displayed in red, and the ERBR model's predictions with the best-fitting parameter, $\lambda \approx 0.69$, are displayed in green.}
		\label{fig:exp2.1}
	\end{figure}

	Figure \ref{fig:exp2.1} compares the true probabilities for each event, the mean reported beliefs for each partition, and the predictions of the ERBR model where the entropy regularization parameter is held constant across all partitions. The analysis in this section assumes that the subjects' benchmark beliefs coincide with the true objective probability distribution. Alternatively, as illustrated in the next section, one can deduce the latent benchmark prior from reported beliefs, and use the latent benchmark prior instead. Since the recovered benchmark prior is reasonably close to the objective probability distribution, this does not significantly impact the results in this section.
	
	The best-fitting parameter for the model ($\lambda \approx 0.69$) was determined by finding the value of $\lambda$ that minimizes the average Root Mean Squared Error (RMSE) between the model's predictions and the empirical means. To prevent the 11 distinct partitions in $\mathcal{P}_{4,k}$ from having a disproportionate impact on the result, this format was treated as a single entity and given equal weight to the other three formats $\mathcal{P}_1$, $\mathcal{P}_2$, and $\mathcal{P}_3$.
	
	The ERBR model with a single entropy regularization parameter significantly improves the fit overall and matches the reported beliefs quite well, particularly for partitions $\mathcal{P}_1$ and $\mathcal{P}_{4,k}$. However, the improvement in model fit is partition-dependent. For $\mathcal{P}_2$, using $\lambda = 1$ (representing true probabilities) provides a better fit than the model, and for $\mathcal{P}_3$, using $\lambda = 0$ (representing uniform probabilities) yields a better fit (see Table \ref{tab:rmse_comparison}). This empirical pattern is in contradiction with the assumption common in recent variational literature that the regularization parameter is partition-invariant and motivates a generalization of the model allowing for partition-dependence. 
	
	\begin{table}[htbp]
		\centering
		\begin{tabular}{l c c c}
			\toprule
			\textbf{Partition} & \textbf{RMSE at $\boldsymbol{\lambda \approx 0.69}$} & \textbf{RMSE at $\boldsymbol{\lambda=0}$} & \textbf{RMSE at $\boldsymbol{\lambda=1}$} \\
			\midrule
			$\mathcal{P}_1$     & \textbf{0.0320} & 0.0746 & 0.0324 \\
			$\mathcal{P}_2$     & 0.0450 & 0.0615 & \textbf{0.0377} \\
			$\mathcal{P}_3$     & 0.0406 & \textbf{0.0048} & 0.0587 \\
			$\mathcal{P}_{4,k}$ & \textbf{0.0265} & 0.3627 & 0.0615 \\
			\bottomrule
		\end{tabular}
		\caption{Root Mean Squared Error (RMSE) comparison across partitions. The lowest RMSE for each partition format is shown in bold.}
		\label{tab:rmse_comparison}
	\end{table}

	\section{Partition-Dependent Entropy Regularization}
	\label{sec:partitiondep}

	In this section, I formalize the extension of the model that allows the entropy regularization parameter to be partition-dependent. As discussed in the Introduction, this extension captures the intuition that an agent’s confidence in her benchmark prior often varies depending on the specific events highlighted by the partition structure. The model is formalized in Definition \ref{def:model2}.

	\begin{definition}
		\label{def:model2}
		A belief reporting rule $\{\mu_{\mathcal{P}}\}_{\mathcal{P}\in\mathbb{P}}$ is consistent with the partition-dependent ERBR model if there exist a benchmark prior $\pi\in \Delta(\Omega)$, assumed to have full support, and entropy regularization parameters $\lambda_{\mathcal{P}}\in \mathbb{R}$ for each $\mathcal{P}\in \mathbb{P}$ such that
		\begin{equation}
			\label{eq:reported_beliefs2}
			\mu_\mathcal{P}(E_i) = \frac{\pi(E_i)^{\lambda_{\mathcal{P}}}}{\sum_{E_j\in \mathcal{P}} \pi(E_j)^{\lambda_{\mathcal{P}}}} .
		\end{equation}
	\end{definition}
	
\begin{figure}[htbp]
		\centering
		\includegraphics[width=\linewidth]{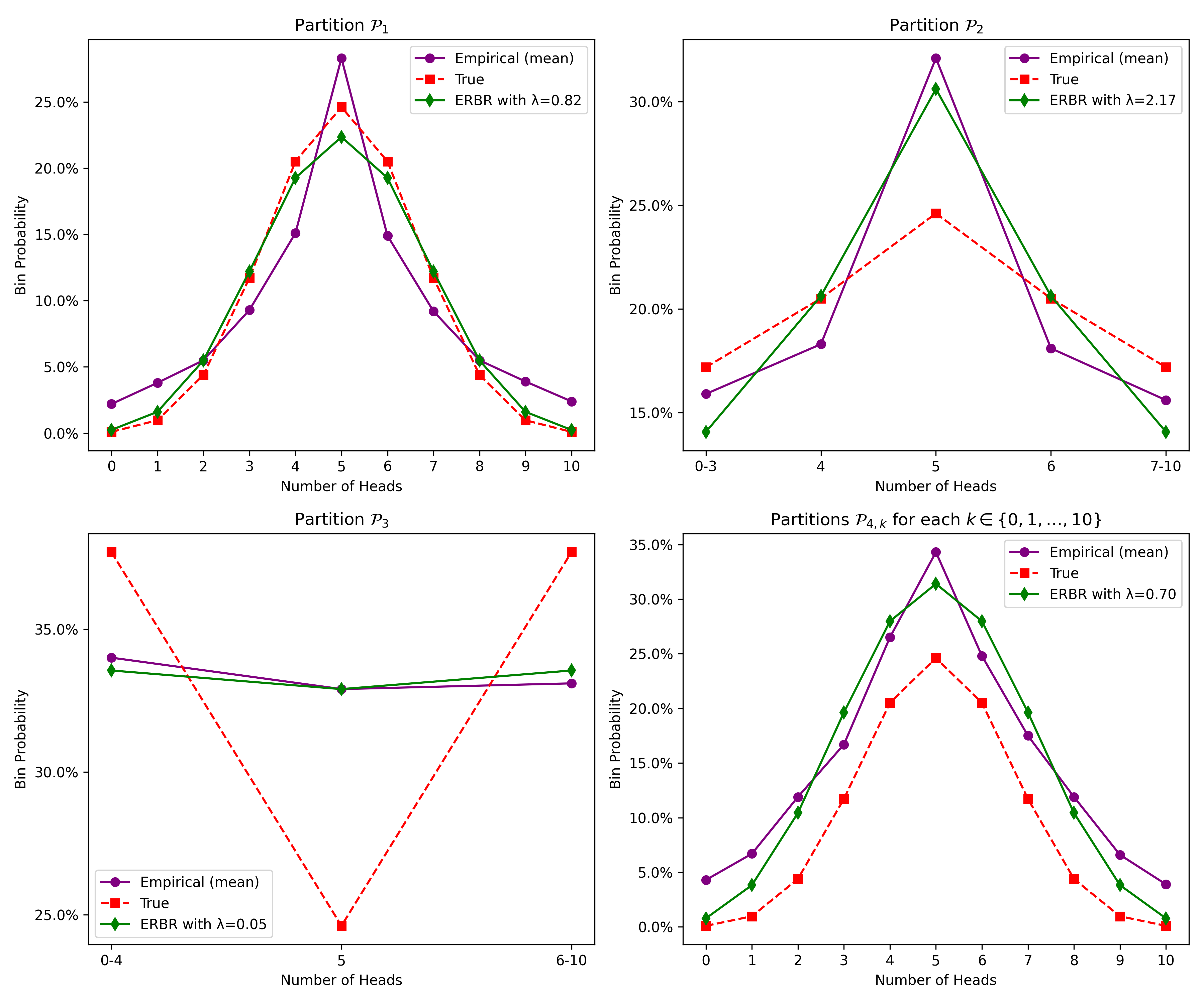} 
		\caption{Model fit to Experiment 2 in \cite{benjamin2017biased} with partition-dependent parameters.} 
		\label{fig:exp2.2} 
	\end{figure}

	Figure \ref{fig:exp2.2} demonstrates the model fit to the experimental data from \citet{benjamin2017biased}, assuming that the subjects' latent benchmark prior matches the true probability distribution. For partitions $\mathcal{P}_{4,k}$, the figure uses the same parameter across all 11 partitions. Note that for binary partitions, one can always find the appropriate $\lambda_{\mathcal{P}}$ to perfectly match the observed data. Figure \ref{fig:exp2.2} shows that even when one restricts the entropy regularization parameter to be constant for all 11 binary partitions in $\mathcal{P}_{4,k}$, the model still offers a significant improvement over the rational benchmark.
	
	As shown in Table  \ref{tab:lambda_comparison}, the additional flexibility of partition-dependent parameters yields relatively minor improvements for partitions $\mathcal{P}_1$ and $\mathcal{P}_{4,k}$, confirming that the restriction to a single parameter adequately captures belief reporting in these partitions. For the other two partitions, there are substantial gains in model fit with partition dependence and the regularization parameters diverge sharply from the original estimate. Specifically, for $\mathcal{P}_2$, the best-fitting $\lambda$ is significantly above one. This suggests that subjects are highly confident in their evaluation that having five heads in ten tosses is much more likely than the alternatives, resulting in an overconfident reporting of beliefs. On the other hand, the parameter for $\mathcal{P}_3$ is quite close to zero. In this case, subjects seem to perceive the event probabilities as being very close to one another, leading them to default to a uniform distribution. Overall, allowing for partition dependence in entropy regularization significantly improves the model fit. 
	
	\begin{table}[htbp]
		\centering
		\begin{tabular}{l c c c}
			\toprule
			\textbf{Partition} & \textbf{Optimal $\boldsymbol{\lambda_{\mathcal{P}}}$} & \textbf{RMSE (Partition-Dep.)} & \textbf{RMSE (Single)} \\
			\midrule
			$\mathcal{P}_1$		& 0.82 & 0.0313 & 0.0320 \\
			$\mathcal{P}_2$      & 2.17 & 0.0198 & 0.0450 \\
			$\mathcal{P}_3$      & 0.05 & 0.0037& 0.0406 \\
			$\mathcal{P}_{4,k}$ & 0.70 & 0.0263 & 0.0265 \\
			\bottomrule
		\end{tabular}
		\caption{Comparison of model fit with a single versus a partition-dependent regularization parameter.}
		\label{tab:lambda_comparison}
	\end{table}
		
	To discipline the model and enable the recovery of latent beliefs, we can impose further structure on how the parameter $\lambda$ changes across partitions. For example, the analysis in Figure \ref{fig:exp2.2} suggests that one can restrict $\lambda$ to be the same for all binary partitions of the type $\mathcal{P}_{\omega} = \{\{\omega\}, \{\omega\}^c\}$ without any significant loss in explanatory power. Imposing this restriction allows the analyst to uniquely recover the latent benchmark prior if reported beliefs are observed for this set of binary partitions. To see this, let $\mu_{\omega} = \mu_{\mathcal{P}_{\omega}}(\{\omega\})$ and assume $\lambda_{\mathcal{P}_{\omega}}=\lambda\neq 0$ for all $\omega\in \Omega$. The model then satisfies 
	$$\mu_{\omega}= \frac{\pi(\omega)^{\lambda}}{\pi(\omega)^{\lambda} + (1-\pi(\omega))^{\lambda}}.$$
	Therefore,
	$$\frac{\pi(\omega)}{1-\pi(\omega)} = \left(\frac{\mu_{\omega}}{1-\mu_{\omega}}\right)^{1/\lambda} \quad \Rightarrow \quad \pi(\omega) = \frac{\left(\frac{\mu_{\omega}}{1-\mu_{\omega}}\right)^{1/\lambda}}{1+\left(\frac{\mu_{\omega}}{1-\mu_{\omega}}\right)^{1/\lambda}}.$$
	Using the fact that probabilities sum to one, we get 
	$$\sum_{\omega\in \Omega} \frac{\left(\frac{\mu_{\omega}}{1-\mu_{\omega}}\right)^{1/\lambda}}{1+\left(\frac{\mu_{\omega}}{1-\mu_{\omega}}\right)^{1/\lambda}} = 1,$$
	which provides a single equation that can be numerically solved for $\lambda$ when a solution exists. We can then use the above expression to find $\pi(\omega)$ for each $\omega\in \Omega$.
	
\begin{figure}[htbp]
		\centering
		\includegraphics[width=0.75\textwidth]{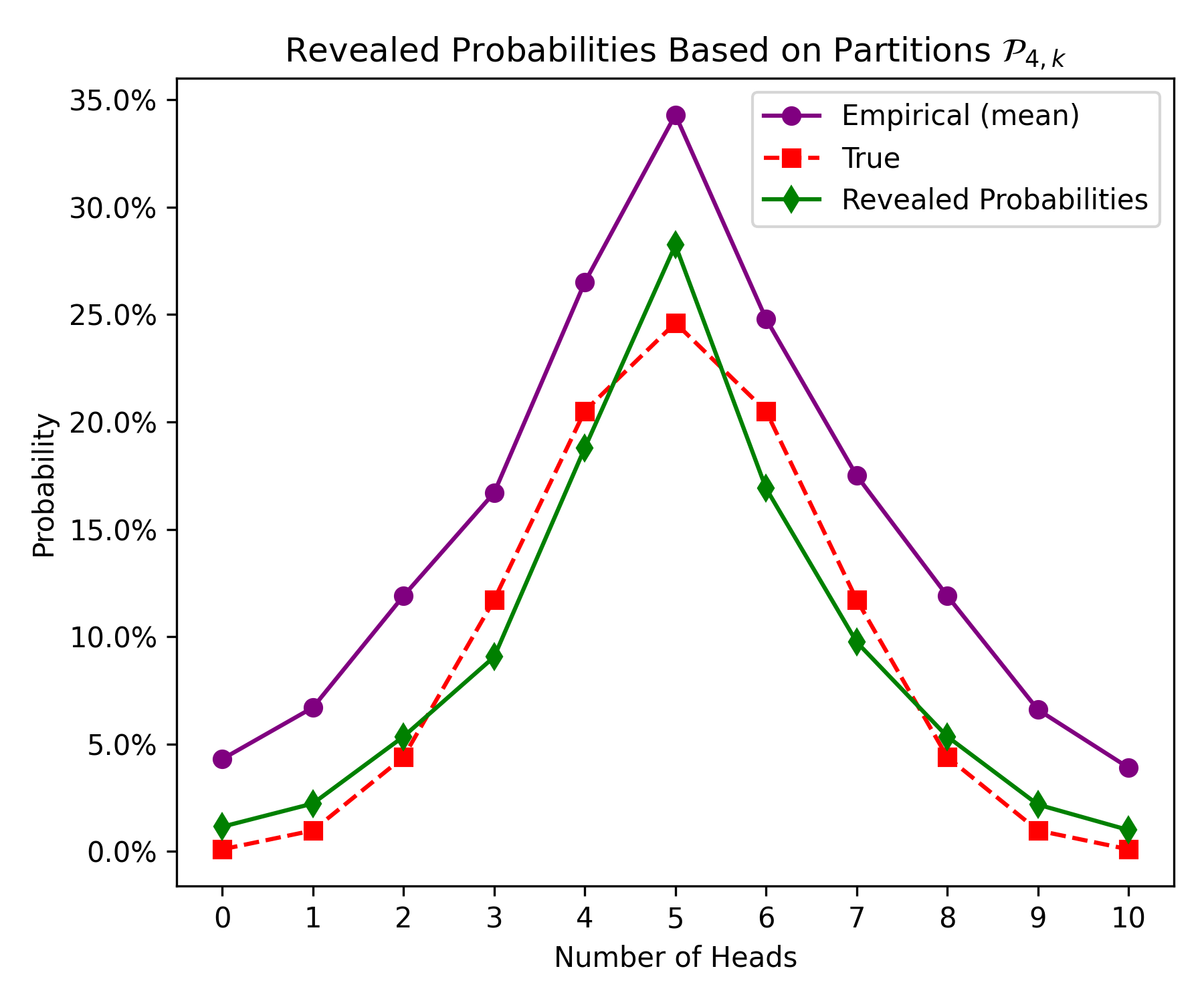}
		\caption{Recovered benchmark probabilities with recovered $\lambda \approx 0.70$. } 
		\label{fig:exp2.3} 
	\end{figure}

	Applying this method to partitions $\mathcal{P}_{4,k}$ in the experimental data recovers the latent benchmark probabilities (Figure \ref{fig:exp2.3}) and a constant $\lambda$ of approximately 0.70.  The recovered benchmark prior aligns closely with the true probabilities, with two notable exceptions: subjects overestimate the likelihood of both the central outcome of five heads and extreme events. These are well-documented biases consistent with past literature \citep{benjamin2017biased}. The fact that the belief recovery method described here generates a latent prior that exhibits these known cognitive biases, which are distinct from partition dependence, supports the use of the ERBR model to disentangle partition effects from underlying cognitive biases. 
	
\begin{figure}[htbp]
		\centering
		\includegraphics[width=\textwidth]{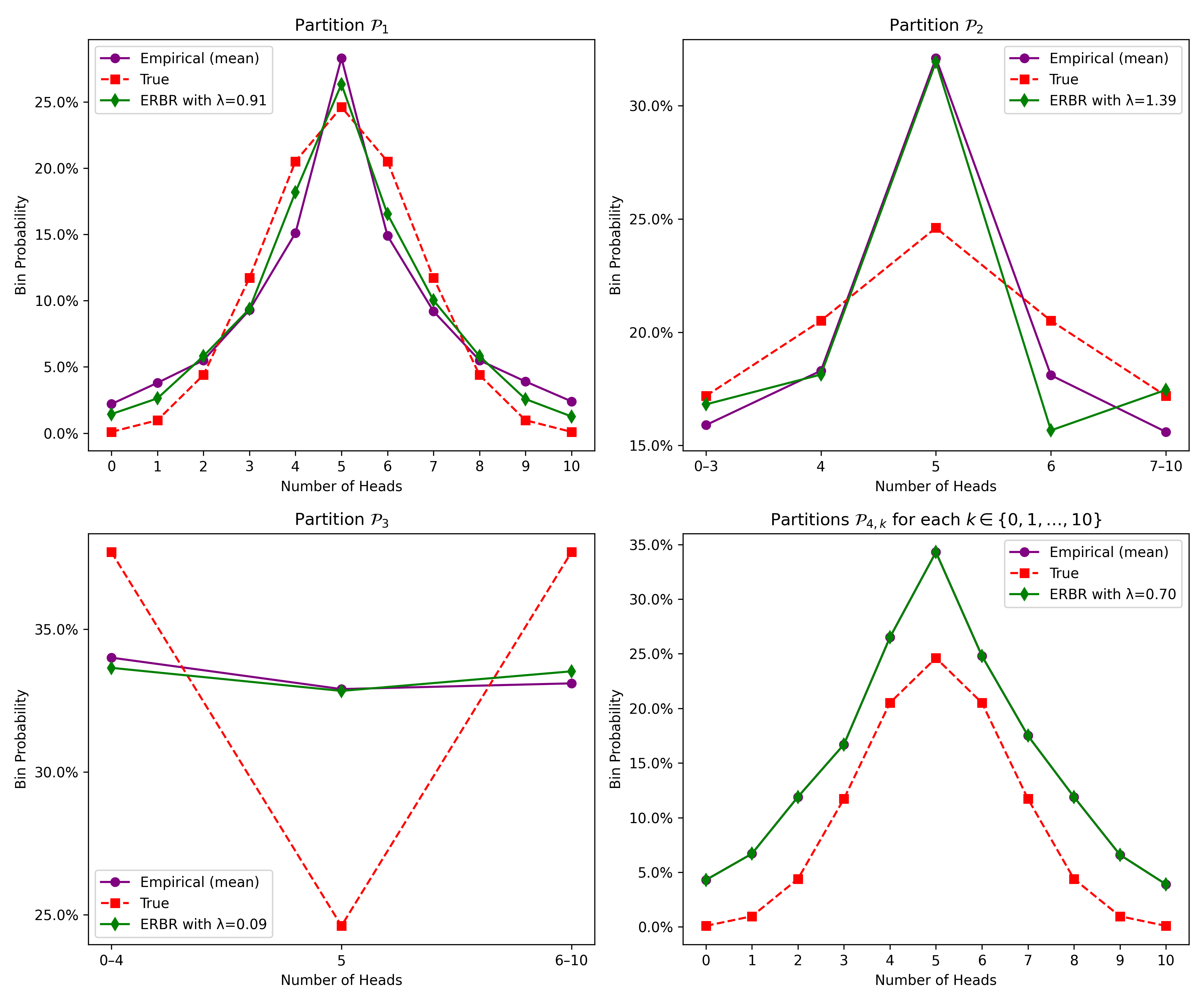}
		\caption{Model fit to Experiment 2 in \cite{benjamin2017biased} with partition-dependent parameters and recovered benchmark probabilities.} 
		\label{fig:exp2.4}
	\end{figure}

	Repeating the analysis from Figure \ref{fig:exp2.2} using the recovered probabilities as the benchmark prior yields the model fit illustrated in Figure \ref{fig:exp2.4}. Table \ref{tab:prior_comparison} compares the results of this approach to those derived using the true probabilities. While the fit for the $\mathcal{P}_{4,k}$ partitions is necessarily perfect, as the prior was recovered from this data, the model using the recovered prior also significantly improves the out-of-sample fit for partitions $\mathcal{P}_1$, $\mathcal{P}_2$, and $\mathcal{P}_3$. In addition, using the recovered prior results in best-fitting $\lambda_{\mathcal{P}}$ values that are closer to one, particularly for $\mathcal{P}_1$ and $\mathcal{P}_2$. This also supports the hypothesis that the recovered prior is a more accurate reflection of subjects' underlying benchmark beliefs than the true probabilities.
	
	\begin{table}[htbp]
		\centering
		\begin{tabular}{l cc cc}
			\toprule
			& \multicolumn{2}{c}{\textbf{Using True Prior}} & \multicolumn{2}{c}{\textbf{Using Recovered Prior}} \\
			\cmidrule(lr){2-3} \cmidrule(lr){4-5}
			\textbf{Partition} & \textbf{Optimal $\boldsymbol{\lambda_{\mathcal{P}}}$} & \textbf{RMSE} & \textbf{Optimal $\boldsymbol{\lambda_{\mathcal{P}}}$} & \textbf{RMSE} \\
			\midrule
			$\mathcal{P}_1$    & 0.82 & 0.0313 & 0.91 & 0.0141 \\
			$\mathcal{P}_2$   & 2.17 & 0.0198 & 1.39 & 0.0143 \\
			$\mathcal{P}_3$   & 0.05 & 0.0037 & 0.09 & 0.0032 \\
			$\mathcal{P}_{4,k}$  & 0.70 & 0.0263 & 0.70 & 0.0000 \\
			\bottomrule
		\end{tabular}
		\caption{Comparison of model fit using true vs. recovered benchmark priors.}
		\label{tab:prior_comparison}
	\end{table}
	
	\cite{benjamin2017biased} consider a general model of partition dependence given by
	\begin{equation}
		\label{eq:root}
		\mu_\mathcal{P}(E_i) = \frac{g_{\mathcal{P}}(\pi(E_i))}{\sum_{E_j\in \mathcal{P}} g_{\mathcal{P}}(\pi(E_j))}.
	\end{equation}
	They assume that the functions $g_{\mathcal{P}}$ are strictly increasing and concave. Under these assumptions, the partition-dependent ERBR model is a special case of equation \eqref{eq:root} where $g_{\mathcal{P}}(x) = x^{\lambda_{\mathcal{P}}}$, provided $\lambda_{\mathcal{P}}\in (0,1]$. While \cite{benjamin2017biased} do not propose a specific functional form for $g_{\mathcal{P}}$, they use this framework to disentangle partition dependence from other biases such as the Law of Small Numbers \citep{kahneman1971belief}, Non-Belief in the Law of Large Numbers \citep{benjamin2016model}, and Exact Representativeness \citep{camerer1987biases}. The analysis in this paper provides a microfoundation for this general framework. The specific functional form of the ERBR model also enables the recovery of the latent prior that exhibits these biases. This confirms the necessity of controlling for partition dependence in belief reporting as argued in \cite{benjamin2017biased}.
	
	\subsection{Conjunction and Disjunction Fallacies} 
	
	The partition-dependent ERBR model can also accommodate widely observed anomalies in belief reporting such as conjunction and disjunction fallacies \citep{tversky1983extensional, barhillel1993how}. The conjunction fallacy refers to the phenomenon that, in certain cases, the conjunction of two events $A$ and $B$, i.e., the event $A\cap B$, is considered more likely than either $A$ or $B$ or both. The disjunction fallacy refers to the phenomenon that the disjunction of two events $A$ and $B$, i.e., the event $A\cup B$, is considered less likely than either $A$ or $B$ or both. In both instances, the agent assigns a smaller probability to a larger event. I mostly focus on the conjunction fallacy, as symmetric arguments can also be used for the disjunction fallacy. 
	
	The most famous version of the conjunction fallacy is known as the Linda problem, reported in \citet{tversky1983extensional}. In their experiment, subjects were provided the following description of a hypothetical person named Linda: 
	
	\begin{quote}
		``Linda is 31 years old, single, outspoken and very bright. She majored in philosophy. As a student, she was deeply concerned with issues of discrimination and social justice, and also participated in anti-nuclear demonstrations.''
	\end{quote} 
	
	The subjects were asked to evaluate the relative likelihood of the following events:
	\begin{enumerate}[(a)]
		\item Linda is a bank teller (event $A$).
		\item Linda is active in the feminist movement (event $B$).
		\item Linda is a bank teller and is active in the feminist movement (event $A\cap B$).
	\end{enumerate}
	
	The vast majority of subjects rank the event $B$ as the most likely and the event $A$ as the least likely, even though the event $A\cap B$ is contained in the event $A$. In their baseline experiment, \citet{tversky1983extensional} report that roughly 85\% of subjects rank $A\cap B$ as more likely than $A$. They discuss various modifications of the experiment that make the relationship $A\cap B\subset A$ more transparent. While this reduces the conjunction fallacy effect, the majority of subjects still rank $A\cap B$ as more likely than $A$. 
	
	While the predominant explanation for the conjunction fallacy emphasizes the representativeness heuristic of \citet{kahneman1972subjective}, the partition-dependent ERBR model can also accommodate this phenomenon as follows.\footnote{The single-parameter ERBR model can also accommodate conjunction and disjunction fallacies. However, the explanation involves assuming $\lambda<0$, which implies a preference for misreporting. This is unlikely to be a satisfactory explanation for these fallacies, as they have been reported across numerous experiments involving a wide variety of subjects.} When agents are asked to evaluate the probabilities for the events $A$, $B$, and $A\cap B$, they form the corresponding partitions $\mathcal{P}_{A} = \{A, A^c\}$, $\mathcal{P}_{B} = \{B, B^c\}$, and $\mathcal{P}_{A\cap B} = \{A\cap B, (A\cap B)^c\}$. For event $A$, they report their belief based on the partition $\mathcal{P}_{A}$, i.e., $\mu_{\mathcal{P}_{A}}(A)$. Similarly for event $B$, they report $\mu_{\mathcal{P}_{B}}(B)$ and for event $A\cap B$, they report $\mu_{\mathcal{P}_{A\cap B}}(A\cap B)$. This reporting rule would be consistent with the experimental data in \citet{teigen1974subjective}, as described in Section \ref{sec:properties}. Under appropriate entropy regularization parameters $\lambda_{\mathcal{P}_{A}}$ and $\lambda_{\mathcal{P}_{A\cap B}}$, we obtain $\mu_{\mathcal{P}_{A\cap B}}(A\cap B)> \mu_{\mathcal{P}_{A}}(A)$, consistent with the conjunction fallacy. The next proposition formalizes the relationship between entropy regularization parameters for the conjunction fallacy to hold.

	\begin{proposition}
		\label{prop:conjunction}
		Suppose the agent reports her beliefs as in the partition-dependent ERBR model. In addition, when the agent is asked to provide a probability estimate for an event $A$, she reports $\mu_{\mathcal{P}_{A}}(A)$, where $\mathcal{P}_{A} = \{A,A^c\}$. Then, for any event $B\subsetneq C$, 
		$$\mu_{\mathcal{P}_{B}}(B)>\mu_{\mathcal{P}_{C}}(C)$$
		if and only if 
		$$\lambda_{\mathcal{P}_{B}}\ln\left(\frac{\pi(B)}{1-\pi(B)}\right) >
		\lambda_{\mathcal{P}_{C}}\ln\left(\frac{\pi(C)}{1-\pi(C)}\right).$$
	\end{proposition}
	
	Consider the case $\pi(B)<\pi(C)<0.5$ so that the logarithmic terms above are negative. Then, the above inequality implies that $\lambda_{\mathcal{P}_{B}}<\lambda_{\mathcal{P}_C}$. Alternatively, if $\pi(C)>\pi(B)>0.5$, we get $\lambda_{\mathcal{P}_{B}}>\lambda_{\mathcal{P}_C}$. The intuition for this is as follows. When $\pi(B)<\pi(C)<0.5$, the event $B$ is considered a more extreme event, and hence the agent applies a stronger degree of entropy regularization for the partition $\mathcal{P}_{B}$ than $\mathcal{P}_{C}$, resulting in $\lambda_{\mathcal{P}_{B}}<\lambda_{\mathcal{P}_C}$. Alternatively, when $\pi(C)>\pi(B)>0.5$, the event $C$ is a more extreme event, and hence the agent applies a stronger degree of entropy regularization for the partition $\mathcal{P}_{C}$. In both cases, the agent is more unsure about the probabilities of extreme events leading to stronger entropy regularization. Interestingly, the model predicts that we are unlikely to observe the conjunction fallacy if $\pi(C)>0.5>\pi(B)$. This is because the above inequality implies that for the conjunction fallacy to hold, either $\lambda_{\mathcal{P}_{B}}$ or $\lambda_{\mathcal{P}_{C}}$ must be negative in this case.
	
	The key mechanism allowing the model to accommodate the conjunction fallacy is that the agent reports beliefs based on the partition $\mathcal{P}_{A\cap B}$ for the event $A\cap B$ and based on the partition $\mathcal{P}_{A}$ for the event $A$. If the agent used the same partition, for example, $\mathcal{P}_{AB}=\{A\setminus B, B\setminus A, A\cap B, (A\cup B)^c\}$, to report beliefs for both events, then the model would not generate the conjunction fallacy. The agent may not use the partition $\mathcal{P}_{AB}$ as it is significantly more complex and involves four distinct events. The partitions $\mathcal{P}_{A}$ and $\mathcal{P}_{A\cap B}$ only involve two events and are much simpler to use to provide estimates for the probabilities of $A$ and $A\cap B$. 
	
	In their analysis, \citet{tversky1983extensional} make a distinction between decompositional and holistic approaches in probability judgments. They argue that for events that follow decompositional logic where the agent divides the event into distinct parts to calculate its probability, we are much less likely to observe the conjunction fallacy. Alternatively, when the probability of an event is evaluated holistically, the conjunction fallacy is more likely to hold. In the framework of the ERBR model, if events follow decompositional logic, then the agent is likely to use a finer partition $\mathcal{P}_{AB}$ to report her beliefs and we are less likely to observe the conjunction fallacy. Alternatively, if the agent uses a holistic approach to evaluate probabilities, then she is likely to use the partitions $\mathcal{P}_{A}$ and $\mathcal{P}_{A\cap B}$, which can lead to reported beliefs that are consistent with the conjunction fallacy. 
	
	\section{Conclusion}
	\label{sec:conclusion}
	
	This paper investigates a tractable model of partition dependence where reported beliefs balance the preference for truthful reporting against the preference for being noncommittal. I provide the structural properties of the model that allow for the recovery of the agent's latent beliefs and apply the model to the experimental data. The analysis shows that the assumption of constant entropy regularization, which is common in recent theoretical literature, is not supported by the data. Extending the model to allow for partition-dependent regularization significantly improves the model fit while maintaining sufficient structure to recover the latent benchmark. 
	
	The ability to recover the latent prior allows the analyst to separate partition dependence from other biases in probabilistic reasoning, as discussed in Section \ref{sec:partitiondep}. For example, one challenge in the literature is separating partition dependence from biases in belief updating. To illustrate, consider the following setup. The analyst first elicits the agent's beliefs over the partition $\mathcal{P}$. Next, the analyst reveals that an event $A\in \sigma(\mathcal{P})$, where $\sigma(\mathcal{P})$ is the algebra generated by $\mathcal{P}$, is realized and elicits the agent's ex post beliefs. The analyst would like to infer if the agent's updating is Bayesian. If the agent performs entropy regularization with a constant $\lambda$ across partitions, then the reported beliefs will satisfy Bayes' rule if and only if the agent updates $\pi$ in a Bayesian manner and reports the entropy-regularized posterior. This is due to the result in \citet{chambers2023coherent}, which identifies a class of belief distortion functions that commute with conditioning. In this case, one could test Bayesian updating by relying on reported beliefs without recovering the latent prior. However, as discussed earlier, the constant parameter assumption is not supported by the data. If the entropy regularization parameter is partition-dependent, the commutativity result no longer holds, and the analyst needs to disentangle partition dependence from biases in belief updating. The partition-dependent ERBR model provides the necessary framework that can be used to recover the latent prior and perform this test.
	
	While the partition-dependent ERBR model significantly improves the explanatory power of the baseline model, it can also be viewed as too flexible. One possible way to discipline the model is by restricting the entropy regularization parameter to be constant for some or all binary partitions, as described in Section \ref{sec:partitiondep}. Alternatively, one can endogenize entropy regularization by linking the entropy regularization parameter to the agent's confidence in her beliefs for each partition. To illustrate, the agent might be endowed with a set of plausible priors $\mathbb{N}$ and the parameter $\lambda_{\mathcal{P}}$ can capture the degree of uncertainty implied by $\mathbb{N}$ for each $\mathcal{P}$. If all plausible priors agree on all events in $\mathcal{P}$, then there is no uncertainty and we can set $\lambda_{\mathcal{P}}=1$. Alternatively, we can let $\lambda_{\mathcal{P}}=0$ when there is maximal uncertainty, with $\lambda_{\mathcal{P}}\in (0,1)$ capturing the intermediate case. An extended model with endogenous parameters would discipline behavior across partitions while retaining enough flexibility to capture observed experimental findings. Developing such an endogenous framework is a key objective for future research.

	\newpage
	
	\section*{Appendix}
	\label{sec:appendix}
	
	\subsection*{Proof of Proposition \ref{prop:exponential}}
	
	The Lagrangian for the optimization problem is 
	$$\lambda \sum_{E_i \in \mathcal{P}}\pi'_{\mathcal{P}}(E_i) \ln\left(\frac{\pi'_{\mathcal{P}}(E_i)}{\pi_{\mathcal{P}}(E_i)}\right) + (1-\lambda)\sum_{E_i \in \mathcal{P}}\pi'_{\mathcal{P}}(E_i) \ln\left(\pi'_{\mathcal{P}}(E_i)\right)+ \eta_{\mathcal{P}} \left(\sum_{E_i\in \mathcal{P}} \pi'_{\mathcal{P}}(E_i)-1\right),$$
	where $\eta_{\mathcal{P}}$ is the multiplier corresponding to the constraint that reported beliefs add up to one. Simplifying, we get
	$$\sum_{E_i \in \mathcal{P}}\pi'_{\mathcal{P}}(E_i) \ln\left(\pi'_{\mathcal{P}}(E_i)\right) -\lambda\sum_{E_i \in \mathcal{P}}\pi'_{\mathcal{P}}(E_i) \ln\left(\pi_{\mathcal{P}}(E_i)\right)+ \eta_{\mathcal{P}} \left(\sum_{E_i\in \mathcal{P}} \pi'_{\mathcal{P}}(E_i)-1\right).$$
	The first-order conditions require
	$$\ln(\pi'_{\mathcal{P}}(E_i)) +1  - \lambda\ln(\pi_{\mathcal{P}}(E_i)) + \eta_{\mathcal{P}} = 0 \quad \Rightarrow \quad 
	\pi'_{\mathcal{P}}(E_i)  = \frac{\pi_{\mathcal{P}}(E_i)^{\lambda}}{\exp(\eta_{\mathcal{P}}+1)} ,$$
	where $\eta_{\mathcal{P}}$ is chosen to satisfy the constraint. Adding over all $E_i\in \mathcal{P}$ and setting it equal to one, we get
	$$\pi'_\mathcal{P}(E_i) = \frac{\pi_\mathcal{P}(E_i)^{\lambda}}{\sum_{E_j\in \mathcal{P}} \pi_\mathcal{P}(E_j)^{\lambda}},$$
	as desired. Since the objective function is strictly convex and the constraint is linear, this uniquely pins down the reported beliefs. 
	
	\subsection*{Proof of Proposition \ref{prop:subadditive}}
	
	Let $p,q\in (0,1)$ be such that $p+q\leq 1$. We want to show that 
	\begin{itemize}
		\item if $\lambda\leq 1$, then $p^{\lambda} +q^{\lambda} \geq (p+q)^{\lambda}$, and
		\item if $\lambda\geq 1$, then $p^{\lambda} +q^{\lambda} \leq (p+q)^{\lambda}$.
	\end{itemize}
	First, consider the case $\lambda<0$. Since $f(x) = x^{\lambda}$ is a decreasing function for $\lambda<0$, $p<p+q$, and $q>0$, we have 
	$$p^{\lambda} +q^{\lambda} \geq p^{\lambda} \geq (p+q)^{\lambda}.$$
	
	Now, suppose $\lambda\geq 0$. If $\lambda=0$, then subadditivity is trivial. Hence, assume $\lambda>0$. Note that the function $f(x) = x^{\lambda}$ is concave for $\lambda \in (0,1]$ and convex for $\lambda\geq 1$. In addition, $f(0) = 0$ for all $\lambda>0$. To conclude the proof, note that since $f(0)=0$, a concave $f$ implies 
	\begin{align*}
		f(p) &= f\left(\frac{p}{p+q}(p+q)\right)\geq \frac{p}{p+q} f(p+q), \text{ and }\\
		f(q) &= f\left(\frac{q}{p+q}(p+q)\right)\geq \frac{q}{p+q} f(p+q).
	\end{align*}
	Hence, $f(p)+f(q)\geq f(p+q)$. Similarly, since $f(0)=0$, a convex $f$ implies superadditivity.
	
	\subsection*{Proof of Proposition \ref{prop:GST}}
	
	The necessity of Regularity follows from the representation in equation \ref{eq:GST} and the assumption that $s(E_i)>0$ for all nonempty $E_i\subseteq \Omega$. The necessity of Cyclical Independence follows from the exposition in the main text.
	
	For sufficiency, we first construct a support function $s(\cdot)$ and then show that it is well-defined. To this end, pick $\omega^*\in \Omega$ and let $s(\{\omega^*\})=1$. For a nonempty $A\subsetneq \Omega$, we can always find a collection of partitions $\mathcal{P}_1, \mathcal{P}_2, \ldots, \mathcal{P}_n$ and a collection of events $E_1,E_2,\ldots, E_{n+1}$ such that $E_1=\{\omega^*\}\in \mathcal{P}_1$, $E_{n+1}=A\in \mathcal{P}_n$, and $E_{i+1}\in \mathcal{P}_i\cap \mathcal{P}_{i+1}$ for $i< n$. We can then define $s(A)$ by
	$$s(A) = \frac{\mu_{\mathcal{P}_n}(E_{n+1})}{\mu_{\mathcal{P}_n}(E_{n})}\frac{\mu_{\mathcal{P}_{n-1}}(E_{n})}{\mu_{\mathcal{P}_{n-1}}(E_{n-1})}\cdots \frac{\mu_{\mathcal{P}_1}(E_2)}{\mu_{\mathcal{P}_1}(E_1)}.$$
	In fact, we can construct $s(A)$ with $n\leq 2$, as illustrated in the main text.

	We need to show that $s(A)$ is well-defined. Suppose there exists another collection of partitions $\mathcal{P}'_1, \mathcal{P}'_2, \ldots, \mathcal{P}'_m$ and events $E'_1,E'_2,\ldots, E'_{m+1}$ such that $E'_1=\{\omega^*\}$, $E'_{m+1}=A$, and $E'_{i+1}\in \mathcal{P}'_i\cap \mathcal{P}'_{i+1}$ for $i<m$. Construct a new collection of partitions $\mathcal{P}''_1, \mathcal{P}''_2, \ldots, \mathcal{P}''_n, \mathcal{P}''_{n+1},\ldots,  \mathcal{P}''_{n+m}$ and events $E''_1, E''_2, \ldots, E''_n, E''_{n+1},\ldots, E''_{n+m}$ by
	$$
	\begin{minipage}{.5\textwidth}
		\[
		\mathcal{P}''_i =
		\begin{cases}
			\mathcal{P}_{n-i+1}, & 1 \le i \le n,\\
			\mathcal{P}'_{i-n},  & n < i \le n+m,
		\end{cases}
		\]
	\end{minipage}%
	\begin{minipage}{.5\textwidth}
		\[
		E''_i =
		\begin{cases}
			E_{n-i+2}, & 1 \le i \le n,\\
			E'_{i-n},  & n< i \le n+m.
		\end{cases}
		\]
	\end{minipage}
	$$
	Notice that $E''_1=A=E_{n+1}=E'_{m+1}$ and $E''_{n+1}=\{\omega^*\}$. In addition, $E''_{j+1}\in \mathcal{P}''_j\cap  \mathcal{P}''_{j+1}$ for $j<n+m$ and $E''_1 \in \mathcal{P}''_{1}\cap \mathcal{P}''_{n+m}$. By Cyclical Independence,
	
	\begin{align*}
		1  & = \frac{\mu_{\mathcal{P}''_1}(E''_1)}{\mu_{\mathcal{P}''_1}(E''_2)}\frac{\mu_{\mathcal{P}''_2}(E''_2)}{\mu_{\mathcal{P}''_2}(E''_3)}\cdots \frac{\mu_{\mathcal{P}''_n}(E''_n)}{\mu_{\mathcal{P}''_n}(E''_{n+1})}\cdot  \frac{\mu_{\mathcal{P}''_{n+1}}(E''_{n+1})}{\mu_{\mathcal{P}''_{n+1}}(E''_{n+2})} \cdots \frac{\mu_{\mathcal{P}''_{n+m}}(E''_{n+m})}{\mu_{\mathcal{P}''_{n+m}}(E''_1)} \\
		& =	\frac{\mu_{\mathcal{P}_n}(E_{n+1})}{\mu_{\mathcal{P}_n}(E_n)}\frac{\mu_{\mathcal{P}_{n-1}}(E_{n})}{\mu_{\mathcal{P}_{n-1}}(E_{n-1})}\cdots \frac{\mu_{\mathcal{P}_1}(E_2)}{\mu_{\mathcal{P}_1}(E_1)}\cdot 
		\frac{\mu_{\mathcal{P}'_1}(E'_1)}{\mu_{\mathcal{P}'_1}(E'_2)}\frac{\mu_{\mathcal{P}'_2}(E'_2)}{\mu_{\mathcal{P}'_2}(E'_3)}\cdots \frac{\mu_{\mathcal{P}'_m}(E'_m)}{\mu_{\mathcal{P}'_m}(E'_{m+1})}\\
		& = s(A)\cdot  \frac{\mu_{\mathcal{P}'_1}(E'_1)}{\mu_{\mathcal{P}'_1}(E'_2)}\frac{\mu_{\mathcal{P}'_2}(E'_2)}{\mu_{\mathcal{P}'_2}(E'_3)}\cdots \frac{\mu_{\mathcal{P}'_m}(E'_m)}{\mu_{\mathcal{P}'_m}(E'_{m+1})},
	\end{align*}
	which implies 
		$$s(A) = \frac{\mu_{\mathcal{P}'_m}(E'_{m+1})}{\mu_{\mathcal{P}'_m}(E'_m)}\frac{\mu_{\mathcal{P}'_{m-1}}(E'_m)}{\mu_{\mathcal{P}'_{m-1}}(E'_{m-1})}\cdots \frac{\mu_{\mathcal{P}'_1}(E'_2)}{\mu_{\mathcal{P}'_1}(E'_1)}.$$
	Hence, $s(A)$ is well-defined for all nonempty $A\subseteq \Omega$ (we can let $s(\Omega)$ be any positive number), and, by Regularity, $s(A)>0$.
	
	To conclude the proof of the representation, we need to show that $\mu_{\mathcal{P}}(\cdot)$ is consistent with the GST model with this choice of support function. Let $\mathcal{P}\in \mathbb{P}$ and $A\in \mathcal{P}$ be given. If $A=\Omega$ and $\mathcal{P}=\{\Omega\}$, there is nothing to show. Alternatively, let $B\in \mathcal{P}$ be distinct from $A$. Then, we can construct a collection of partitions $\mathcal{P}_1, \mathcal{P}_2, \ldots, \mathcal{P}_n$ with $\mathcal{P}_n=\mathcal{P}$ and a collection of events $E_1,E_2,\ldots, E_n, E_{n+1}$ such that $E_1=\{\omega^*\}$, $E_n=A$, $E_{n+1}=B$,  and $E_{i+1}\in \mathcal{P}_i\cap \mathcal{P}_{i+1}$ for $i<n$. Using the definition of $s(\cdot)$,
	\begin{align*}
		s(B) = \frac{\mu_{\mathcal{P}_n}(E_{n+1})}{\mu_{\mathcal{P}_n}(E_n)}\frac{\mu_{\mathcal{P}_{n-1}}(E_n)}{\mu_{\mathcal{P}_{n-1}}(E_{n-1})}\cdots \frac{\mu_{\mathcal{P}_1}(E_2)}{\mu_{\mathcal{P}_1}(E_1)}
		= \frac{\mu_{\mathcal{P}}(B)}{\mu_{\mathcal{P}}(A)} s(A).
	\end{align*}
	Adding over $B\in \mathcal{P}$,
		\begin{align*}
		\sum_{B\in \mathcal{P}} s(B) = \frac{\sum_{B\in \mathcal{P}}\mu_{\mathcal{P}}(B)}{\mu_{\mathcal{P}}(A)} s(A) = \frac{s(A)}{\mu_{\mathcal{P}}(A)},
	\end{align*}
	where $\sum_{B\in \mathcal{P}}\mu_{\mathcal{P}}(B)=1$ due to Regularity. Hence,
	$$\mu_{\mathcal{P}}(A)= \frac{s(A)}{\sum_{B\in \mathcal{P}} s(B)}.$$
	Since this is true for any $\mathcal{P}\in \mathbb{P}$ and $A\in \mathcal{P}$, the representation follows. For uniqueness, note that once we fix $s(\{\omega^*\})$, $s(\cdot)$ is uniquely defined for all other proper events. 
	
	\subsection*{Proof of Proposition \ref{prop:ERBR}}
	
	The necessity of Power Additivity is described in the main text. For sufficiency, note that we already have the GST representation 
	$$\mu_{\mathcal{P}}(A)= \frac{s(A)}{\sum_{B\in \mathcal{P}} s(B)}.$$
	Since $s(\cdot)^{\alpha}$ is additive for some $\alpha\neq 0$, we can let 
	$$\pi(\omega) = \frac{s(\{\omega\})^{\alpha}}{\sum_{\omega'\in \Omega}s(\{\omega'\})^{\alpha}}.$$
	Note that $\pi(\omega)>0$ and $\sum_{\omega\in \Omega}\pi(\omega)=1$, and hence $\pi(\cdot)$ is a full support probability measure. In addition, for $A\subsetneq \Omega$,
	$$\pi(A) = \sum_{\omega\in A} \pi(\omega) 
	=  \frac{\sum_{\omega\in A} s(\{\omega\})^{\alpha}}{\sum_{\omega'\in \Omega}s(\{\omega'\})^{\alpha}} 
	= \frac{ s(A)^{\alpha}}{\sum_{\omega'\in \Omega}s(\{\omega'\})^{\alpha}},$$
	where the last equation is due to Power Additivity. Therefore, 
	$$\pi(A)^{1/\alpha} =  \frac{ s(A)}{\left(\sum_{\omega'\in \Omega}s(\{\omega'\})^{\alpha}\right)^{1/\alpha}} \quad \Rightarrow \quad s(A) = \pi(A)^{1/\alpha} \left(\sum_{\omega'\in \Omega}s(\{\omega'\})^{\alpha}\right)^{1/\alpha} $$
	Combining the previous expressions, 
	\begin{align*}
		\mu_{\mathcal{P}}(A)= \frac{s(A)}{\sum_{B\in \mathcal{P}} s(B)} 
		= \frac{\pi(A)^{1/\alpha} \left(\sum_{\omega'\in \Omega}s(\{\omega'\})^{\alpha}\right)^{1/\alpha}}{ \sum_{B\in \mathcal{P}}\pi(B)^{1/\alpha} \left(\sum_{\omega'\in \Omega}s(\{\omega'\})^{\alpha}\right)^{1/\alpha}} = \frac{\pi(A)^{1/\alpha}}{ \sum_{B\in \mathcal{P}}\pi(B)^{1/\alpha}}.
	\end{align*}
	Hence, letting $\lambda = 1/\alpha$ yields the ERBR representation. 
	
	For uniqueness, if $\{\mu_{\mathcal{P}}\}_{\mathcal{P}\in \mathbb{P}}$ is always uniform, we can identify that $\lambda=0$ but the benchmark prior $\pi$ cannot be recovered. If $\{\mu_{\mathcal{P}}\}_{\mathcal{P}\in \mathbb{P}}$ is not always uniform, then the support function $s(\cdot)$ identified from $\{\mu_{\mathcal{P}}\}_{\mathcal{P}\in \mathbb{P}}$ will not be constant and there will be a unique $\alpha$ satisfying Power Additivity. Letting $\lambda=1/\alpha$ and defining $\pi$ as above uniquely identifies both in this case.
	
	\subsection*{Proof of Proposition \ref{prop:conjunction}}
	
	We have 
	$$\mu_{\mathcal{P}_B}(B)
	=\frac{\pi(B)^{\lambda_{\mathcal{P}_B}}}
	{\pi(B)^{\lambda_{\mathcal{P}_B}} + (1-\pi(B))^{\lambda_{\mathcal{P}_B}}}
	\quad \text{and} \quad 
	\mu_{\mathcal{P}_C}(C)
	=\frac{\pi(C)^{\lambda_{\mathcal{P}_C}}}
	{\pi(C)^{\lambda_{\mathcal{P}_C}} + (1-\pi(C))^{\lambda_{\mathcal{P}_C}}}.$$
	Hence,
	$$\mu_{\mathcal{P}_B}(B)>\mu_{\mathcal{P}_C}(C)
	\quad\Longleftrightarrow\quad
	\frac{\pi(B)^{\lambda_{\mathcal{P}_B}}}
	{\pi(B)^{\lambda_{\mathcal{P}_B}} + (1-\pi(B))^{\lambda_{\mathcal{P}_B}}}
	>
	\frac{\pi(C)^{\lambda_{\mathcal{P}_C}}}
	{\pi(C)^{\lambda_{\mathcal{P}_C}} + (1-\pi(C))^{\lambda_{\mathcal{P}_C}}}.$$
	Since the denominators are positive, cross-multiplying, we get
	$$\pi(B)^{\lambda_{\mathcal{P}_B}}
	\left(\pi(C)^{\lambda_{\mathcal{P}_C}} + (1-\pi(C))^{\lambda_{\mathcal{P}_C}}\right)
	>
	\pi(C)^{\lambda_{\mathcal{P}_C}}
	\left(\pi(B)^{\lambda_{\mathcal{P}_B}} + (1-\pi(B))^{\lambda_{\mathcal{P}_B}}\right).$$
	Expanding and canceling the common term $\pi(B)^{\lambda_{\mathcal{P}_B}}\pi(C)^{\lambda_{\mathcal{P}_C}}$ yields
	$$\pi(B)^{\lambda_{\mathcal{P}_B}} (1-\pi(C))^{\lambda_{\mathcal{P}_C}}
	>
	\pi(C)^{\lambda_{\mathcal{P}_C}} (1-\pi(B))^{\lambda_{\mathcal{P}_B}}.$$
	Dividing both sides by $(1-\pi(B))^{\lambda_{\mathcal{P}_B}}\,(1-\pi(C))^{\lambda_{\mathcal{P}_C}}$, we get
	$$\left(\frac{\pi(B)}{1-\pi(B)}\right)^{\lambda_{\mathcal{P}_B}}
	>
	\left(\frac{\pi(C)}{1-\pi(C)}\right)^{\lambda_{\mathcal{P}_C}}.$$
	Taking the natural logarithms on both sides, we get
	$$\lambda_{\mathcal{P}_B}\ln\left(\frac{\pi(B)}{1-\pi(B)}\right)
	>
	\lambda_{\mathcal{P}_C}\ln\left(\frac{\pi(C)}{1-\pi(C)}\right),$$
	as desired.

\newpage 
	
\bibliographystyle{abbrvnat}
\bibliography{partition}
	
\end{document}